# Optimizing Discrete Multi-tone Transmission for 400G Data Center Interconnects

Annika Dochhan[1], Helmut Griesser[2], Nicklas Eiselt[1,3], Michael Eiselt[1], Jörg-Peter Elbers[2]

(1) ADVA Optical Networking SE, Maerzenquelle 1-3, 98617 Meiningen, Germany

(2) ADVA Optical Networking SE, Fraunhoferstr. 9a, 82152 Martinsried / Munich, Germany

(3) Technical University of Denmark (DTU), Department of Photonics Engineering, Ørsteds Plads, Build. 343, DK-2800

adochhan@advaoptical.com

## Kurzfassung

Discrete Multi-tone Transmission (DMT) ist ein vielversprechender Kandidat für zukünftige 400G Verbindungen zwischen Datenzentren. Acht Kanäle mit jeweils einer Datenrate von 56 Gb/s können in einem 50-GHz-Kanalraster zu einem 400 Gb/s „Superkanal" zusammengefasst werden. Für ein voll beladenes 96-Kanal-DWDM-System führt dies zu einer Gesamtkapazität von 4.8 Tb/s. Um die Anforderungen der angestrebten Anwendung in Form von Kosteneffizienz und geringer Leistungsaufnahme zu erfüllen, ist es wichtig, die Komplexität der digitalen Signalverarbeitung so gering wie möglich zu halten. Für DMT wird die Komplexität hauptsächlich durch die Länge der schnellen Fouriertransformation (FFT) bestimmt. Da Datenzentrenverbindungen typischerweise eine Distanz von bis zu 80 km zu überbrücken haben, wird hier neben anderen Parametern der Einfluss der FFT-Länge auf die Leistungsfähigkeit des Systems mit 56 Gb/s DMT für diese Streckenlänge untersucht. Die Übertragung findet im C-Band statt, um DWDM zu ermöglichen und es wird auf optische Dispersionskompensation verzichtet. Sowohl Zweiseitenband- (DSB) als auch Restseitenband- (VSB) DMT werden betrachtet. Es zeigt sich, dass eine FFT-Länge von 128 ausreichend ist, um die angestrebte Leistungsfähigkeit in Form einer Ziel-Bitfehlerquote zu erreichen, obgleich eine größere Länge die Leistungsfähigkeit signifikant zu steigern vermag.

## Abstract

Discrete multi-tone transmission (DMT) is a promising candidate for future 400G data center interconnects. Eight channels, each carrying 56 Gb/s of data can be combined in a 50-GHz channel grid to form a 400 Gb/s superchannel. For a fully loaded 96 channel DWDM system this leads to a total capacity of 4.8 Tb/s. To meet the requirements of the targeted application in terms of cost efficiency and low power consumption, it is important to keep the complexity for the digital signal processing as low as possible. For DMT, the complexity is mainly determined by the length of the fast Fourier transformation (FFT). Since data center interconnects only have to bridge a distance of typically 80 km, we here investigate among other parameters the influence of the FFT length among other parameters on the achievable performance for 56 Gb/s DMT only for this distance. Transmission is performed in C-band to enable DWDM and no optical dispersion compensation is applied. We consider double sideband (DSB) as well as vestigial sideband (VSB) DMT. It can be seen that an FFT length of 128 is sufficient to reach the required performance in terms of bit error ratio, however, a higher length can significantly improve the performance.

## 1 Introduction

To meet the ever increasing demand of data throughput for inter- and intra-data center connectivity, while keeping cost and complexity low, several advanced intensity modulated and directly detected modulation formats such as discrete multi-tone transmission (DMT), four level pulse amplitude modulation (PAM4) or carrierless amplitude/phase modulation (CAP) have been proposed [1-3]. To mitigate the impact of chromatic dispersion (CD) on the DMT signal, part of the signal spectrum can be filtered out, leading to a vestigial sideband signal instead of a double sideband signal [4]. Recently, vestigial sideband or single sideband DMT has been shown for even higher data rates than 56 Gb/s, e.g. for 100 Gb/s and 80 km reach in [5] and [6], but at the expense of two electrical driving signals and an IQ modulator at the transmit side. Low complexity VSB with detuned filters for 100 Gb/s and 100 km reach enabled by non-linearity compensation was shown in [7], but with broad optical filters. In [8], we proposed a four- to eight-channel DMT system as a possible solution to transmit 400 Gb/s over up to 240 km.

Here, we focus on the eight-channel system with 56 Gb/s on each channel. For this system, previously a reach of 240 km was achieved. This indicates a lot of margin which leaves room for reducing complexity and overhead of the DMT digital signal processing, which is mainly determined by the number of subcarriers (i.e. FFT length). In our previous work, we have chosen the parameters to maximize the performance for the given components, whereas for real-time implementation the complexity should be as low as possible. Here, as a first step, we con-sider a single-channel 56 Gb/s system, but with 50-GHz multiplexer and demultiplexer filter. These filters also perform the vestigial sideband filtering, enabled by a detuned laser frequency. We evaluate the required FFT length, the optimum cyclic prefix and the optimum number of training symbols in case of 80 km transmission distance. It can be seen that an FFT length of 128 is sufficient if the available OSNR is above 32 dB.

## 2  DMT System Setup

**Fig. 1** shows the DMT system setup. The signal was generated offline using python routines and loaded to a high-speed digital-to-analog-converter (DAC) at a sampling rate of 84 GS/s. To optimally adjust the signal spectrum to the transmission channel, bit and power loading (BL, PL) were applied. Chow's margin-adaptive bit loading algorithm [9] and Cioffi's power loading [10] were used to distribute the bits and to adjust the power of the subcarriers according to the previously estimated signal-to-noise ratio (SNR). The SNR estimation was done by evaluating the quality of a detected signal carrying 16-QAM with equal power on all subcarriers. Since a real valued baseband signal is required, the number of usable subcarriers is half of the FFT length minus one. Hence, a 1024-point FFT leads to 511 usable subcarriers. Moreover, an oversampling factor of 1.05 (to support anti-aliasing filtering) leads to a further reduction of data-carrying subcarriers, in case of the 1024-point FFT to 486. To meet the memory requirements of the DAC, a DMT frame consists of 128 symbols. Among these are up to 20 training symbols (TS) for channel estimation and synchronization. If the number of training symbols were increased, the number of payload symbols would be reduced, requiring a higher total data rate for a net rate of 56 Gb/s. The same applies for the cyclic prefix (CP). It is varied between 2 and 512 samples, however, the optimum depends on the amount of chromatic dispersion. The DMT time signal was symmetrically clipped with a clipping ratio of 9 dB to reduce the peak-to-average power ratio. The clipping level was optimized together with the Mach-Zehnder Modulator (MZM) bias voltage to achieve a minimum OSNR requirement for a bit error ratio (BER) of $10^{-3}$.

The DMT signal was modulated onto a continuous wave optical carrier at 194.25 THz, using a dual-drive MZM (suitable for 40 Gb/s signals) and the differential outputs of the DAC. The optimum driving condition is achieved for linear modulation of the electrical field. The signal was then fed into a 96-channel DWDM multiplexer filter with 39 GHz bandwidth per channel. After amplification by a booster Erbium-doped fiber amplifier (EDFA), the input power to the fiber was adjusted by a variable optical attenuator (VOA). After transmission over 80 km of standard single mode fiber (SSMF) the signal was amplified by a second EDFA, preceded by a VOA to perform noise loading, if necessary. The EDFA was followed by an optical demultiplexer with the same characteristics as the multiplexer and by another VOA to adjust the input power into the receiver. The receiver consisted of a 30-GHz bandwidth PIN photodiode in combination with a transimpedance amplifier (TIA). In contrast to our previous setup [4, 8], where only a PIN was used, we here avoid the need of an additional EDFA at the receiver side, making the setup more realistic for practical application. In addition, the PIN-TIA was followed by an 84 GS/s analog-to-digital converter (ADC) and not by a high-performance real-time oscilloscope. At this point, it has to be emphasized that the output swing of the TIA was less than half of the ADC input dynamic. In a practical implementation, the TIA output would be perfectly matched to the ADC, but for the experiment no other PIN-TIA was available. Thus, a slightly better noise performance of the system than presented here might be possible. However, the parameter optimization process is independent of this. The detected signals were processed offline in python routines, as follows: a Schmidl-Cox [11] like synchronization was applied, followed by the FFT and the removal of CP and TS. Then, a one-tap equalization with decision-directed channel estimation was performed. After demapping of the symbols, the BER was calculated. For all experiments we consider hard decision forward error correction (FEC) with a BER limit of $3.8 \cdot 10^{-3}$.

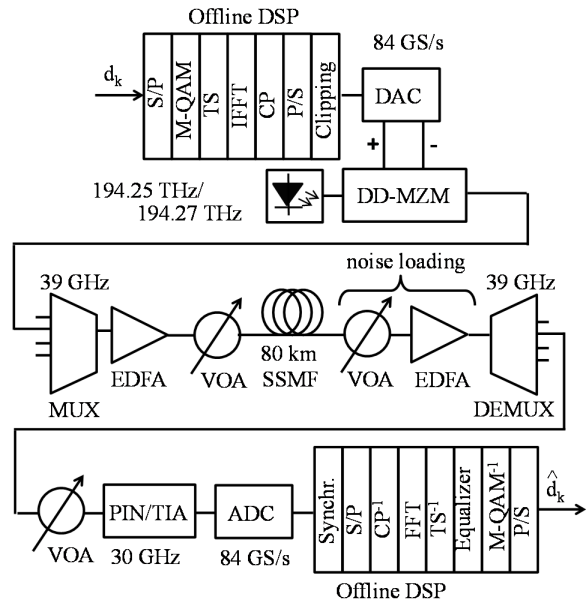

**Figure 1** System Setup. Offline generated DMT frames were transmitted over 80 km SSMF with 39-GHz MUX and DEMUX at TX and RX side, respectively.

## 3  Results and Discussion

### 3.1  Launch Power

The optimization of the system is intended for 80 km transmission. In a first step, it is necessary to determine the optimum input power into the fiber, which should be independent of the other parameters. Therefore, for a first evaluation, a DSB DMT signal with an FFT length of 512 and a CP of 8 samples was chosen. The launch power into the fiber was varied between -7 and +9 dBm, showing an optimum at 5 dBm, which is in perfect agreement with [4]. Later on, for the optimized VSB signal (FFT length 512, 32 samples CP) this optimum launch power was verified. Both results are depicted in **Fig. 2**.

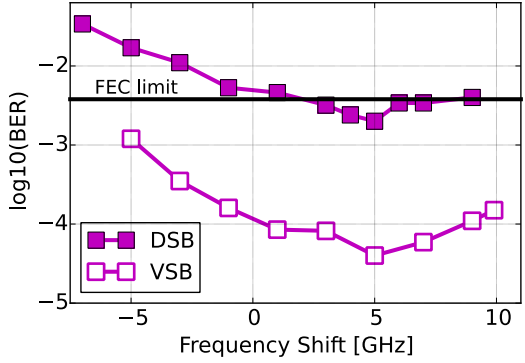

**Figure 2** BER vs. launch power for an FFT length of 512. DSB signal with 4 samples CP, VSB signal optimized, with 32 samples CP. Optimum launch power is 5 dBm.

## 3.2 Cyclic Prefix

With 5 dBm input power, we now varied the FFT length and the length of the CP. Theoretically the CP can be calculated by [12], [13]

$$T_{CP} \geq D \cdot L \cdot \frac{c}{f^2} \cdot B_{DMT}, \quad (1)$$

where $D$ is the dispersion coefficient of the fiber, $L$ its length, $c$ the speed of light, $f$ the carrier frequency and $B_{DMT}$ the bandwidth of the DMT signal. If we assume a bandwidth of 40 GHz for the signal, the required number of samples for the CP can be calculated as approximately 36. A too long cyclic prefix would lead to a higher data rate to keep the net rate at 56 Gb/s and thus reduce the performance.

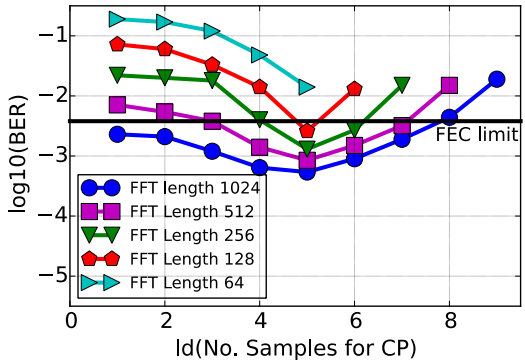

**Figure 3** BER vs. ld(CP samples) for DSB DMT and 80 km transmission. Optimum for all FFT lengths is 5, corresponding to 32 CP samples.

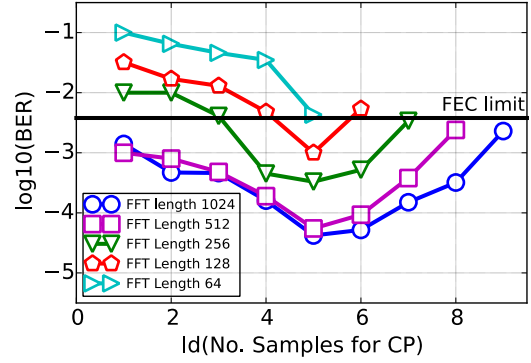

**Figure 4** BER vs. ld(CP samples) for VSB DMT and 80 km transmission. Optimum for all FFT lengths is 5, corresponding to 32 CP samples.

**Fig. 3** shows the achievable BER versus the dual logarithm (ld) of the number of CP samples. The optimum is 32 samples and is very close to the calculated value. The results for the VSB case are shown in Fig. 4. Again 32 CP samples are the optimum. The VSB filtering reduces the influence of dispersion notches (power fading), but the inter-symbol interference (ISI) stays the same. It can be seen that for an FFT length of 64 the FEC limit cannot be reached in any case. Therefore, this length will not be considered in the following.

## 3.3 Number of Training Symbols

Next, the number of training symbols (TS) was evaluated. Due to memory requirements in terms of page length and total size of the DAC and the ADC and to keep the reading and writing speed low, a maximum frame length of 128 symbols is considered. We varied the number of TS inside this frame at the expense of the payload data. To keep the net data rate constant, a higher number of TS leads to a higher total bit rate. The results are shown in **Fig. 5**. The optimum number of TS lies between 4 and 5. We chose 5 for all further experiments.

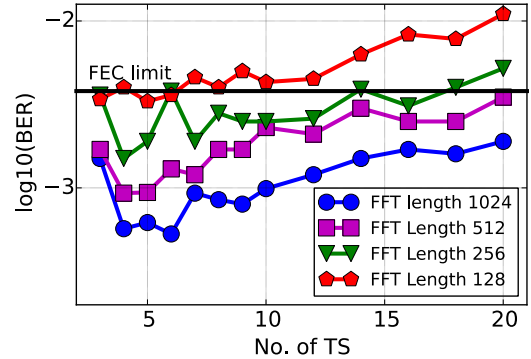

**Figure 5** BER vs. number of training symbols for DSB DMT, 80 km transmission and different FFT lengths. Total number of symbols per frame (payload + TS) is 128 in all cases.

## 3.4 Frequency Shift for VSB

Although we already presented VSB results in the previous sections of this paper, the optimization of the relative shift between signal carrier and passband center of the multiplexer and demultiplexer was not shown. This evaluation was performed with the 128-point FFT signal, with the optimum CP of 32 samples. **Fig. 6** shows the BER vs. the frequency deviation between passband center and signal carrier. The minimum BER was achieved for a detuning of 20 GHz. Therefore, for all VSB experiments the laser frequency was detuned by 20 GHz, leading to a carrier at 194.27 GHz.

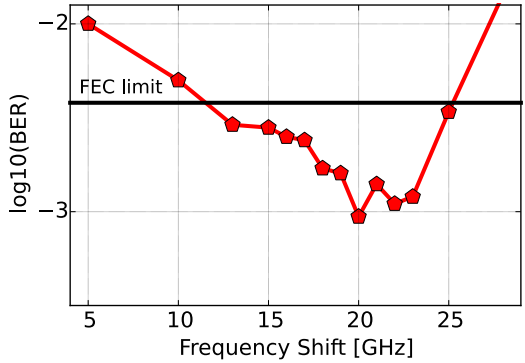

**Figure 6** BER vs. frequency shift for 80 km transmission, FFT length 128, and a CP of 32 samples.

## 3.5 Transmission Result

As a last step, the transmission performance vs. optical noise for the FFT lengths of 1024, 512, 256 and 128 with the optimum CP of 32 samples was evaluated. **Fig. 7** to **10** show the BER vs. optical SNR (OSNR) for the back-to-back case (b2b), as dashed curves and for 80 km transmission as solid lines. The DSB results are presented with filled markers, whereas the unfilled markers indicate the VSB results.

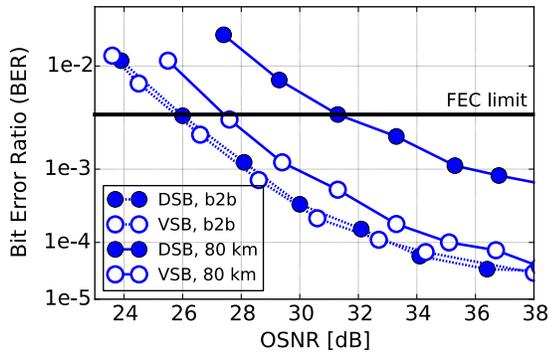

**Figure 7** FFT length 1024: BER vs. OSNR for b2b and 80 km transmission, for both, DSB and VSB.

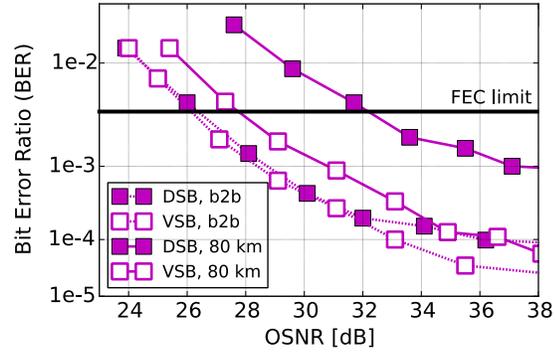

**Figure 8** FFT length 512: BER vs. OSNR for b2b and 80 km transmission, for both, DSB and VSB.

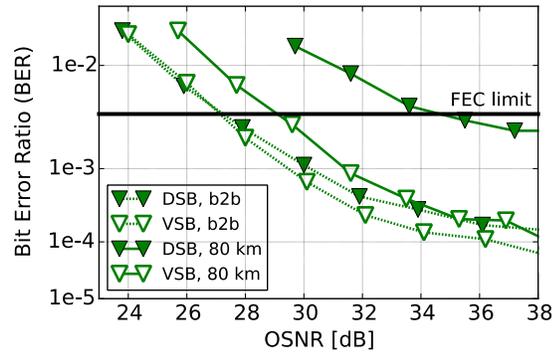

**Figure 9** FFT length 256: BER vs. OSNR for b2b and 80 km transmission, for both, DSB and VSB.

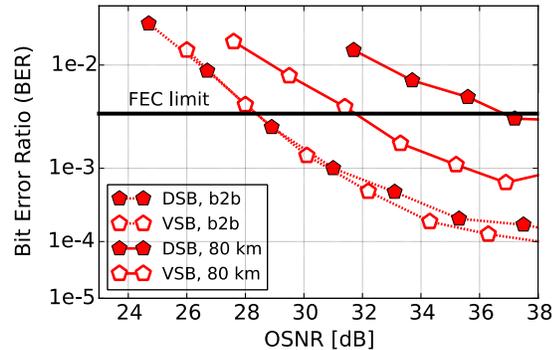

**Figure 10** FFT length 128: BER vs. OSNR for b2b and 80 km transmission, for both, DSB and VSB.

In all cases, the b2b results show similar performance for DSB and VSB. In fact, the VSB results are slightly better than the DSB results. The 39-GHz optical filtering imposes a bandwidth limitation on the signal. In case of central filtering, part of the spectrum is lost and the other subcarriers need to carry more data. However, if the signal is asymmetrically filtered, information for the high frequency subcarriers can be recovered from the sideband which is completely inside the passband. This leads to an increased usable bandwidth. This effect can clearly been seen in **Fig. 11** and **12**, where the estimated SNR spectra for the b2b case for FFT lengths of 1024 and 128 are depicted.

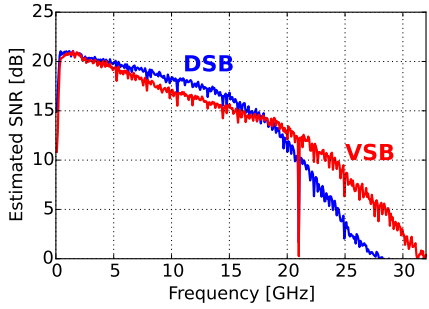

**Figure 11** FFT length 1024: Estimated SNR for DSB and VSB, b2b.

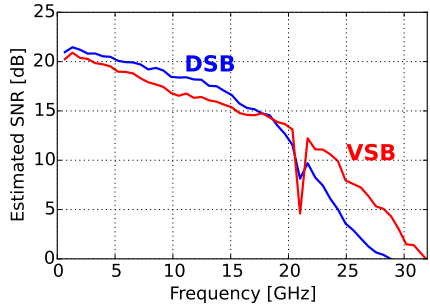

**Figure 12** FFT length 128: Estimated SNR for DSB and VSB, b2b.

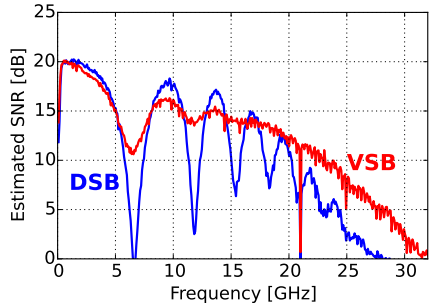

**Figure 13** FFT length 1024: Estimated SNR for DSB and VSB, 80 km.

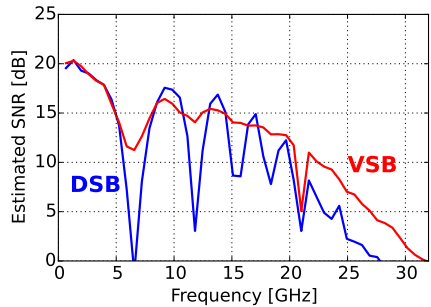

**Figure 14** FFT length 128: Estimated SNR for DSB and VSB, 80 km.

For all FFT lengths, DSB transmission over 80 km leads to a strong signal degradation with at least 5 dB of OSNR penalty due to the dispersion induced notches in the SNR spectrum. These can be seen in **Fig. 13** and **14** for the DSB case, again for FFT lengths 1024 and 128. Thanks to VSB filtering, this penalty can be reduced by 4 dB or more.

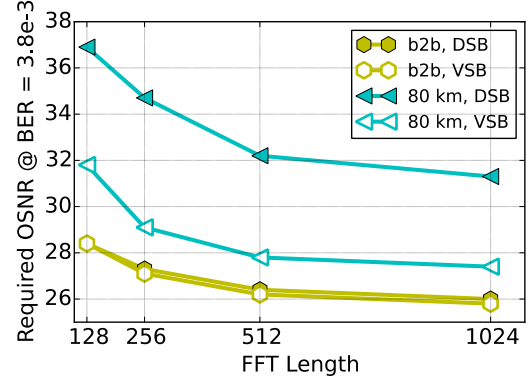

**Figure 15** Required OSNR at a BER of $3.8 \cdot 10^{-3}$ vs. FFT length for b2b and 80 km transmission and for DSB and VSB.

To allow a better comparison between the different FFT lengths, **Fig. 15** shows the required OSNR for BER of $3.8 \cdot 10^{-3}$ vs. FFT length for b2b and 80 km, for both DSB and VSB. A longer FFT length enables a more precise adaption of the signal to the channel characteristics, thus a longer FFT improves the system. However, increasing the FFT length further from 512 to 1024, the performance improvement flattens out. It can be concluded that for this system, especially if VSB is performed, the use of an FFT length of 512 is a good performance/complexity trade off. However, to keep the complexity as low as possible, an FFT length of 128 still achieves the FEC limit, even in the DSB case.

## 4  Conclusion

We evaluated the performance of a 56-Gb/s 80-km DMT transmission system, using both double and vestigial sideband transmission. An FFT length of 512 turns out to be the best compromise between complexity and performance. However, even a length of 128 can be sufficient with some degradations. A transmission distance of 80 km requires 32 samples of cyclic prefix at the sampling rate of 84 GHz.

## 5  Acknowledgement

The results were obtained in the framework of the SASER ADVAntage-NET and SpeeD projects, partly funded by the German ministry of education and research (BMBF) under contracts 16BP12400 and 13N13744, and by the European Commission in the Marie Curie project ABACUS.